\documentclass[%
reprint,
groupedaddress,
 amsmath,amssymb,
prb,
]{revtex4-2}

\usepackage{amsmath}
\usepackage{graphicx}
\usepackage{epstopdf}
\DeclareGraphicsExtensions{.eps,.pdf,.png,.jpg,.gif}
\usepackage{dcolumn}
\usepackage{color}
\usepackage{xcolor}
\usepackage{float}
\usepackage{bm}

\newcommand{\vect}[1]{\mathbf{#1}}
\newcommand{\mat}[1]{\mathbf{#1}}
\def\RV{\textcolor{black}}
\def\LG{\textcolor{black}}

\begin{document}


\title{Energy Transfer and Coherence in Coupled Oscillators with Delayed Coupling: \\ A Classical Picture \LG{for} Two-Level Systems}

\author{Fahhad H Alharbi}
\email{fahhad.alharbi@kfupm.edu.sa}
\affiliation{Electrical Engineering Department, King Fahd University of Petroleum and Minerals (KFUPM), Dhahran, Saudi Arabia}
\affiliation{SDAIA-KFUPM Joint Research Center for Artificial Intelligence, Dhahran, Saudi Arabia}

\author{Abdelrahman S Abdelrahman}
\affiliation{Electrical Engineering Department, King Fahd University of Petroleum and Minerals (KFUPM), Dhahran, Saudi Arabia}

\author{Abdullah M Alkathiry}
\affiliation{Electrical Engineering Department, King Fahd University of Petroleum and Minerals (KFUPM), Dhahran, Saudi Arabia}

\author{Hussain M Al-Qahtani}
\affiliation{Mechanical Engineering Department, King Fahd University of Petroleum and Minerals (KFUPM), Dhahran, Saudi Arabia}

\date{\today}

\begin{abstract}

The Frimmer-Novotny model to simulate two-level systems by coupled oscillators is extended by incorporating a constant time delay in the coupling. The effects of the introduced delay on system dynamics and two-level modeling are then investigated and found substantial. Mathematically, introducing a delay converts the dynamical system from a finite  one  into an infinite-dimensional system. The \LG{resulted system of delay differential equations} is solved using the Krylov method with Chebyshev interpolation and post-processing refinement. The calculations and analyses \LG{reveal} the critical role that a delay can play. It has oscillatory effects as the main dynamical eigenmodes move around a circle with a radius proportional to the coupling strength and an angle linear with the delay. This alteration governs the energy transfer dynamics and coherence. \RV{Accordingly, both, the delay and the coupling strength dictate the stability of the system. The delay is the main related parameter as for certain intervals of it, the system remains stable regardless of the coupling.} A significant effect occurs when one of the main modes crosses the imaginary axis, where it becomes pure imaginary and dampingless. Thus, the two states energies can live and be exchanged for an extremely long time. Furthermore, it is found that the delay alters both the splitting and the linewidth in a way further influencing the energy transfer and coherence. \RV{It is found also that the delay should not be large to have significant effect. For example, for an optical system with 500 nm wavelength, the critical delay can be in tens of attoseconds.}

\end{abstract}

\maketitle


\section{\label{SecInt}Introduction}
\LG{Coupled oscillators are presumably the most commonly used "building blocks" in modeling} classical and quantum physical phenomena. \cite{D01,F01,H01,N01,R01,AD01,AD02}.
\LG{They are also used as "physical means" to make some ideas and concepts that are difficult to put into practice,} like general-purpose quantum computing \cite{C01} and nonlocal complex system synchronization \cite{S01}. Furthermore, and \LG{perhaps} more intriguingly, coupled oscillators are used to synthesize \LG{a wide range} of seemingly unrelated physical phenomena. \LG{Several quantum phenomena that were originally believed to be purely quantum mechanical in nature, are successfully modeled by classical coupledharmonic oscillators (CHOs).} \LG{Examples include} quantum mechanical two-level system (TLS) \cite{F03,F05}, rapid adiabatic passage \cite{S02}, electromagnetically induced transparency (EIT) \cite{G01}, Stückelberg interferometry \cite{I01,F04}, quantum band formation \cite{R02}, quantum coherence \cite{L01}, and energy transfer \cite{F06,W01}. \LG{We refer the reader to some excellent reviews for information on recent developments.} \cite{C01,K03,I01,R04}.

Conventionally, it is assumed that the coupling interaction is instantaneous. However, with the emergence of attosecond experimental capabilities, there is a a growing interest in \LG{introducing} time delay \LG{into} the coupling \cite{S03,K01,H02,O01,H04}. Actually, this was even proceeded by considering
coupling with induced delay\LG{, as} in optical resonators \cite{K02,L02}, non-local coupled oscillators \cite{R05,S04,P01,Y01}, and systems with delay feedback \cite{S05,C03,A02,H03}. In this work, we \LG{extend} the Frimmer-Novotny model \RV{to emulate two-level systems} \cite{F03,F05,R03} by incorporating a constant time delay \LG{into} the coupling and \LG{studying} the delay effects on system dynamics. \LG{When a delay is introduced, the dynamical system is transformed} from a finite to an infinite-dimensional system \cite{O02,F07,A01}. \LG{Briggs and Eisfeld demonstrated} that the quantum and classical energy transfer and coherence are equivalent \LG{in instantaneous cases \cite{B06} and subsequently} used classical coupled oscillators for quantum dynamics simulations \cite{B07}.

\LG{By including the delay, a system of delay differential equations is created, which is then solved by linearizing} around the instantaneous case using the Krylov method with Chebyshev interpolation and post-processing refinement \cite{J01,G02} . \LG{The model is then used to analyze the dynamics of a two-level system that exhibits delayed interaction. Remarkably, the delay has a significant effect on the dynamics of ultrafast phenomena, and thus provides a suitable foundation for explaining it.} This paper is \LG{intended to be broad in scope and more symbolic in nature, avoiding any particular} systems or range of values. \LG{Nevertheless, with appropriate dimension scaling and transition, the results can be directly translated to practical systems.}

The calculations and analyses \LG{carried out here} illustrate the critical role that a delay can play. \LG{When it comes to system stability, the coupling strength is the most destabilizing factor. However, even with substantial coupling, the delay can help to alleviate this instability and thus stabilize the system.} \LG{Additionally, it was found that the delay has oscillatory effects,} as the main dynamical eigenmodes evolve around a circle with a radius proportional to the coupling strength and an angle proportional to the delay.. Accordingly, the spectrum of these modes can be described very accurately using a simple empirical form. \LG{Under certain conditions,} damping in at least one of the main dynamical modes is \LG{eliminated, allowing a significant portion of the energy to survive} for an extremely long time. \LG{In other instances,} the frequency difference between the two \LG{primary} dynamical modes can be \LG{canceled in such a way that energy exchange is prevented. In the case of} the two-level system, it is \LG{observed that delay has an effect on} both the splitting and the linewidth (a manifestation of damping cancellation). \LG{As a result, the energy transfer and coherence effects are altered.}  

With \LG{the introduced} delay, the studied \LG{system} can \LG{become} unstable when the coupling is strong. However, \LG{within} certain ranges of the delay $\tau$, the system \LG{remains} stable. \LG{When} $\tau=0$, the system is globally stable and remains so until a critical delay, $\tau_{Cri}$\LG{, is reached, at which point} the system becomes dampingless and the energies continue to live and exchange for an extremely long period of time. The \LG{required} delay should not be large and and it
can be manifested in \LG{real-world} systems. \LG{For instance}, $\tau_{Cri}$ can be on the order of tens of attoseconds in an optical system with a 500 nm wavelength.

In the next section, the model will be presented alongside the resulting governing delay differential equations. \LG{In addition, the specifics of solving these delay differential equations, including eigenpairs calculations and stability analyses, are discussed in detail.} The resulting eigenpairs are then used to \LG{determine} the dynamics. \LG{Afterwards, we'll go over the metrics employed to gauge} energy transfer and coherence in systems.  \LG{Results and discussion are provided in the third section of this paper. A comparison of delays and no delays is made. As a starting point for further interpretation,} the instantaneous case is used as a baseline. \LG{Finally, a summary of the findings concludes the paper.}

\section{\label{SecFor} Coupled Oscillators with Time Delay}

\subsection{The Model}

\begin{figure}[ht]
\centering
\includegraphics[width=3.2in]{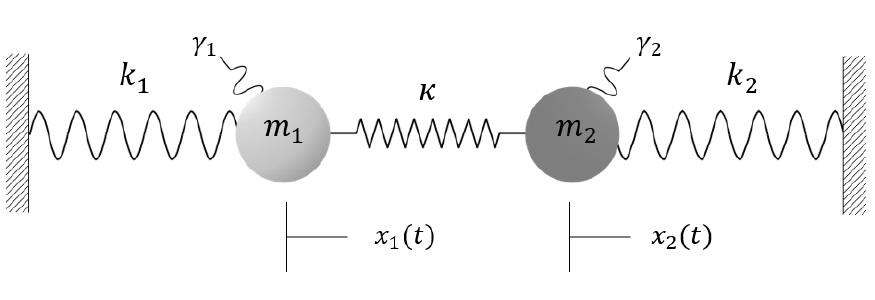}
\caption{The considered mass-spring system model.}
\label{Mdl1}
\end{figure}  

The considered model \LG{as} depicted in Fig. \ref{Mdl1} is composed of two coupled oscillators with masses $m_1$ and $m_2$ and spring \LG{with} constants $k_1$ and $k_2$. The coupling \LG{force} is assumed to be \LG{Hookean using} a spring with a coupling constant $\kappa$. However, we assume that the coupling is not instantaneous and hence the coupling force \RV{on the $i^\text{th}$ mass due to the $j^\text{th}$ mass} takes the \LG{following} form\LG{:}
\begin{equation}
    F_{ij} = \kappa \left[ x_j(t-\tau) -x_i(t) \right]\,. 
\end{equation}
\RV{where the delay can be attributed to a finite interaction speed.} Furthermore, it is assumed that \LG{the} oscillators are damped at rates $\gamma_1$ and $\gamma_2$, respectively and that there are no external driving forces. Thus, the system is described by the following equations of motion:
\begin{equation}
    \label{CEqM}
    \begin{split}
        & \ddot x_1(t) + \gamma_1 \dot x_1(t) + \frac{k_1 + \kappa}{m_1} x_1(t) - \frac{\kappa}{m_1} x_2(t-\tau) = 0, \\
        & \ddot x_2(t) + \gamma_2 \dot x_2(t) + \frac{k_2 + \kappa}{m_2} x_2(t) - \frac{\kappa}{m_2} x_1(t-\tau) = 0.
    \end{split}
\end{equation}

\LG{This coupled problem is best solved} in a state-space representation \cite{R03}, where the following state variables are used:
\begin{equation}
    \begin{split}
        & y_1 = x_1, \\
        & y_2 = \dot x_1, \\
        & y_3 = x_2, \\
        & y_4 = \dot x_2.
    \end{split}
    \label{yvec}
\end{equation}
So, Eq. (\ref{CEqM}) becomes
\begin{equation}
    \label{SSRep}
    \frac{d\vect{y}(t)}{dt}=\mat{A}\,\vect{y}(t)+\mat{B}\,\vect{y}(t-\tau)\, ,
\end{equation}
where $\vect{y}(t)$ is a column vector combining $y_i$'s and the matrices $\mat{A}$ and $\mat{B}$ are
\begin{equation}
    \mat{A} = 
    \begin{bmatrix}
        0 & 1 & 0 & 0 \\
        -\frac{k_1 + \kappa}{m_1} & -\gamma_1 & 0 & 0 \\
        0 & 0 & 0 & 1 \\
        0 & 0 & -\frac{k_2 + \kappa}{m_2} & -\gamma_2
    \end{bmatrix},
\end{equation}
\begin{equation}
    \mat{B} = 
    \begin{bmatrix}
        0 & 0 & 0 & 0 \\
        0 & 0 & \frac{\kappa}{m_1} & 0  \\
        0 & 0 & 0 & 0 \\
        \frac{\kappa}{m_2} & 0 & 0 & 0
    \end{bmatrix}.
\end{equation}

\subsection{Eigenpair Calculations}

Eq. (\ref{SSRep}) is a linear autonomous delay differential equation {LDDE}\LG{, which} is rigorously studied by mathematicians \cite{H05,L05,B02}. Unlike instantaneous differential equations where existence and uniqueness are the main mathematical concerns, LDDE and delay differential equations -- in general -- \LG{require consideration of} solution smoothness as well. \LG{A full history function $\boldsymbol{\phi}(t)$  rather than a set of points serve as the initial conditions}, where:
\begin{equation}
    \label{HistRep}
    \vect{y}(t)=\boldsymbol{\phi}(t), \quad \quad -\tau \leq t \leq 0 \, .
\end{equation}
Giving that $\boldsymbol{\phi}(t)$ is $C^n$ continuous on $[-\tau, \,0]$, then, it is proved that there exists a unique  $C^{n+1}$ continuous solution $\vect{y}(t)$ on $[0, \, \infty)$. Mathematically, Eqs. (\ref{SSRep}) and (\ref{HistRep}) combined constitute the full LDDE system such that:
\begin{equation}
    \label{LDDEs}
    \begin{cases}
      \frac{d\vect{y}(t)}{dt}=\mat{A}\,\vect{y}(t)+\mat{B}\,\vect{y}(t-\tau)\, , & t\geq 0 \\
      \vect{y}(t)=\boldsymbol{\phi}(t)\, , & -\tau \leq t \leq 0 \,.
    \end{cases}
\end{equation}

For delay differential equations, finding the solution relies -- in most cases -- on converting the problem from being finite-dimensional delay differential equations into infinite-dimensional ordinary differential equations by the method of steps \cite{B03} or by the evolution operator approach devised by Krasnosel\'skii \cite{L05,K04,H05}. Favorably, LDDE is among the simplest delay differential equations \LG{and it is analytically solvable. The} solution is simply given by:
\begin{equation}
\label{EPSol}
    \vect{y}(t) = \sum_{k=1}^\infty \sum_{l_k=0}^{m_k-1} c_{k,l_k} t^{l_k} e^{\lambda_k t} \vect{v}_{k,l_k}, \quad t \geq 0
\end{equation}
where $\vect{v}_{k,l_k}$ is the generalized eigenvectors for the $k^\text{th}$ eigenvalue of multiplicity $m_k$. The expansion coefficients $c_{k,l_k}$ are obtained by satisfying the history function $\boldsymbol{\phi}(t)$ for $t\in[-\tau, \, 0]$. If there is no degeneracy for any of the eigenvalues (i.e. all $m_k$ equal 1), then all the vector polynomials become simply constant eigenvectors and the solution is reduced to 
\begin{equation}
    \label{EPSol0}
    \mathbf{y}(t) = \sum_{k=1}^\infty c_k e^{\lambda_k t} \mathbf{v}_k, \quad t \geq 0.
\end{equation}
This is actually the case in the \LG{problem under consideration} due to coupling, as $\kappa \neq 0$.

The eigenspace $\{ \lambda_k, \vect{v}_{k,l_k} \}$ is of infinite-dimension as there is an infinite number of eigenpairs due to the time delay $\tau$. Hale and Lunel \cite{H05,L05} show that $\{ \lambda_k, \vect{v}_k(t) \}$ is simply the null space of 
\begin{equation}
    \label{NSpace}
    \left( \lambda \mat{I} - \mat{A} - \mat{B} e^{-\lambda_k \tau} \right) \vect{v}_k(t) = 0 \, .
\end{equation}
For the current problem (Eq. (\ref{LDDEs})), it is found that \cite{L05, F08}:
\begin{itemize}
    \item all the eigenvalues lie in the complex half-plane $\Re(\lambda_k) < \alpha$ for some $\alpha \in \mathbb{R}$, 
    \item the real parts of eigenvalues (i.e. $\Re(\lambda_k)$) accumulate at $-\infty$,
    \item there is only a finite number of eigenvalues at any vertical strip of the complex plane,
    \item the spacing between eigenvalues decreases with $\tau$.
\end{itemize}

Due to the accumulation of the real parts of the eigenvalues at $-\infty$, most of them have extreme negative real parts, i.e. they decay very rapidly \RV{and thus only have a small impact on the dynamics of the system over a short period of time.} \LG{Hence}, we will focus \LG{-- here --} on the $M$ most-right eigenpairs (i.e. a total of $M$ eigenpairs with the largest real parts of the eigenvalues). This \LG{approach} is valid in the current system as these most-right pairs naturally dominate the dynamics over a relatively large time span. \LG{To locate the $M$ most-right eigenpairs, we employ the Krylov method with Chebyshev interpolation.}\cite{J01,G02,A03}. The basic idea of the method follows the Krasnosel\'skii evolution operator approach, which exploits the fact that any linear delay differential problem (as Eq. (\ref{LDDEs})) can be represented equivalently by a linear infinite-dimensional ordinary differential operator. Then, Chebyshev interpolation is used to approximate the infinite-dimensional operator by a finite one over an interval of time. The subsequent approximation \LG{generate} errors that \LG{depend} on $\tau$ and the used Chebyshev nodes.\LG{The resulting eigenpairs are subsequently}  corrected by post-processing to ensure that the eigenpairs satisfy Eq. (\ref{NSpace}).

\subsection{Solution Stability}
\label{SubSecSt}

The stability of the considered problem requires ensuring contractivity, boundedness, and asymptotic stability \cite{B03,B04,B05}. \LG{In general, the system's stability is $\tau$-dependent, however, this dependency is eliminated under certain conditions.} Theoretically, most stability analyses and conditions are derived from Razumikhin-type theorems on dynamical stability \cite{M01}, Lyapunov exponents \cite{G03}, and spectral analysis \cite{L06}. 
The following conditions are found to be sufficient for stability:
\begin{itemize}
    \item $\Re (\lambda_k) < 0$ (i.e. $\alpha<0$),
    \item $\mu \left[ \mat{A} \right] + \| \mat{B} \| < 0$, and
    \item $\sup_{\Re (\zeta)=0} \rho \left[ (\zeta \mat{I}-\mat{A})^{-1} \mat{B} \right] < 1$
\end{itemize}
where $\mu \left[ \, \cdot \, \right]$ is the logarithmic norm and $\| \cdot \|$ is a norm, and $\rho\left[ \, \cdot \, \right]$ is the spectral radius. \LG{It's worth noting that all these three conditions are correlated.}. The logarithmic norm \cite{S06} is loosely named ``norm'' as it allows negative \LG{values}; $\mu \left[ \mat{A} \right]$ must be negative to satisfy the second condition. In the present work, we assume the maximum norm. 

The first condition implies that all the real parts of the eigenvalues must be negative to ensure asymptotic stability as $t \to \infty$. \LG{For the second condition to be satisfied, the following is required:}
\begin{equation}
    \mu \left[ \mat{A} \right] = \sup_{\vect{u \neq 0}} \frac{\Re \left< \vect{u} | \mat{A} \vect{u} \right>}{\left< \vect{u} | \vect{u} \right>} = \frac{-1}{2}\max (\gamma_1, \, \gamma_2) \, .
\end{equation}
This is due to the fact that the eigenvalues of $\mat{A}$ are $\frac{-\gamma_j}{2} \pm \frac{1}{2} \sqrt{ \gamma_j^2 - 4 \frac{k_j + \kappa}{m_j}}$ where $j=1, 2$. \LG{As a result, the second condition of stability becomes}
\begin{equation}
    \label{StCon2}
    - \max \left( \frac{\gamma_1}{2}, \, \frac{\gamma_2}{2} \right) +
    \max \left( \frac{\kappa}{m_1}, \, \frac{\kappa}{m_2} \right) < 0.
\end{equation}
The last condition necessitates that $\frac{\kappa^2}{(k_1+\kappa)(k_2+\kappa)}<1$ which is globally satisfied. For identical coupled oscillators, Eq. (\ref{StCon2}) is reduced to requiring that $\kappa/m<\gamma/2$ for a stable system. 


\begin{figure*}[ht]
\centering
\includegraphics[width=7.0in]{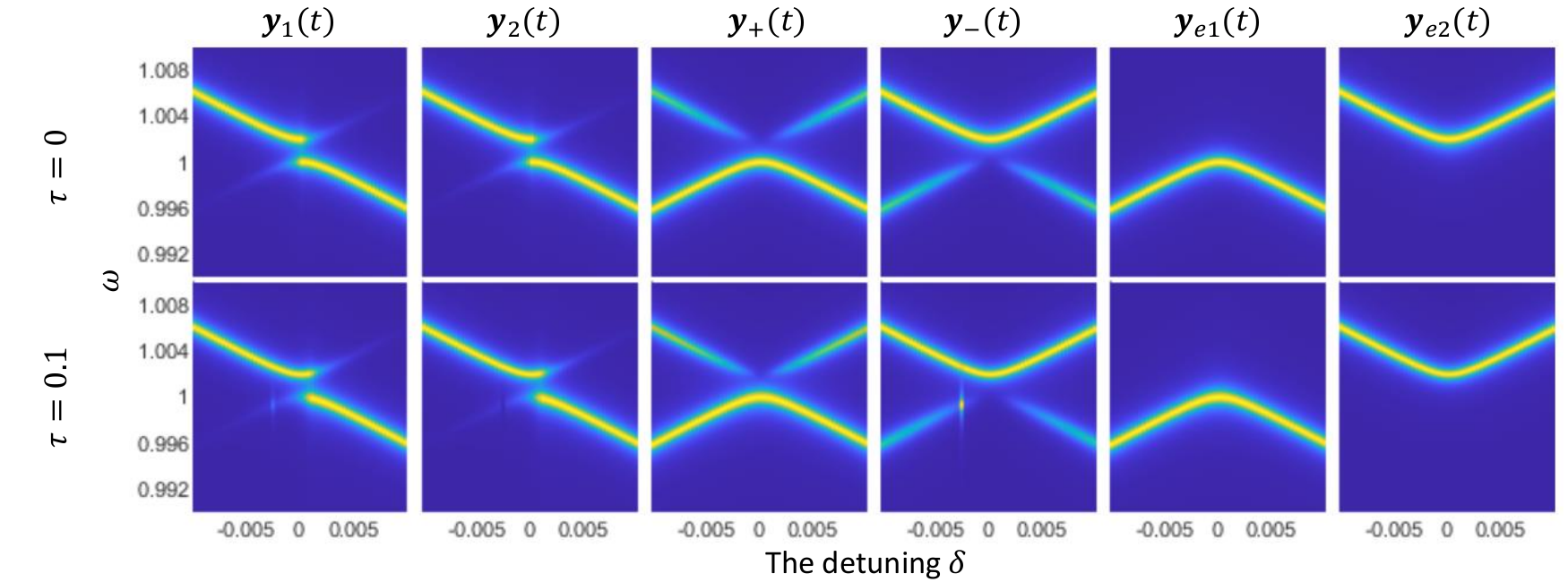}
\caption{The total power spectral density $P_t(\omega)$ vs. a detuning parameter $\delta$ for constant history functions $\vect{y}_{1},\, \vect{y}_{2},\, \vect{y}_{+},\, \& \, \vect{y}_{-}$ besides $\vect{y}_{e1} \, \& \, \vect{y}_{e2}$ corresponding to the lower and upper eigenstates respectively. The upper panels are for the cases with no delay ($\tau=0$) while the lower panels are for $\tau=0.1$.}
\label{TLSFig}
\end{figure*}

\subsection{Two-Level System and its Dynamics}

So far, the used representation is a direct dynamical state-space one. To model a two-level system \LG{by} the considered coupled oscillators, one needs to connect the obtained dynamics to the desired state quantities and then transform the representation into eigenmodes of these quantities \cite{L03,F03,I01,R03}. \LG{For the purposes of this study, we are primarily concerned with two systems that operate at different energies.} So, we need to obtain the \LG{corresponding} energy spectrum, which is proportional to the square of Fourier transforms of the oscillators' dynamics. By using the $M$ most-right eigenpairs, the transforms can be calculated directly from Eq. (\ref{EPSol}) (along with a part -- usually small -- arises from the history function) as follows:
\begin{equation}
    \label{FSSol}
    \vect{Y}(\omega) =
    \sum_{k=1}^\infty \sum_{l_k=0}^{m_k-1} c_{k,l_k} \frac{l_k!}{(i \omega - \lambda_k)^{l_k-1}} \vect{v}_{k,l_k}
\end{equation}
Thus, the power spectral density of each oscillator according to the used state-space representation in Eq. (\ref{yvec}) becomes
\begin{equation}
\begin{split}
    & P_1(\omega) = \frac{\omega_1}{(2\pi)^2} \left( \frac{k_1+\kappa}{2} |\vect{Y}_1(\omega)|^2 + \frac{m_1}{2} |\vect{Y}_2(\omega)|^2 \right) \\
    & P_2(\omega) = \frac{\omega_2}{(2\pi)^2} \left( \frac{k_2+\kappa}{2} |\vect{Y}_3(\omega)|^2 + \frac{m_2}{2} |\vect{Y}_4(\omega)|^2 \right)
\end{split}
\label{PowDen}
\end{equation}
where $\omega_j^2=(k_j+\kappa)/m_j$ \cite{R03}. For each oscillator, the contributions from the potential and kinetic energies are added up. Frimmer and Novotny used a single frequency spectral point as $x_j(t)$ is assumed to take the form $x_j(t)=X_{\Omega_0} \exp (i \Omega_j t)$. The two approaches are equivalent as the full power density  (i.e. Eq. (\ref{PowDen})) is concentrated around the oscillators' normal frequencies $\omega_j$'s with some linewidths. However, assuming that $\gamma_j=0$ -- as in the case of Frimmer and Novotny model --, the linewidth approaches 0 and hence we end up with an infinitesimally sharp spectral line.

The total power spectral density of the system $P_t(\omega)=P_1(\omega)+P_2(\omega)$ represents the desired state \LG{quantity}; but not as eigenmodes. \LG{The two-oscillator system under consideration has} $P_t(\omega)$ concentrated around the oscillators' normal frequencies\LG{, hence it} resembles a two-level system. The corresponding eigenmodes ($\vect{v}_{k,l_k}$) are obtained \LG{through} eigenspace calculation and  are directly related to the dynamical state-space $\vect{y}(t_0)$ at some particular time $t_0$ by direct transformation \cite{F03}
\begin{equation}
    \vect{v}_{k,l_k} = \mat{U}_{k,l_k} \vect{y}(t_0) \, .
\end{equation}
For the two-level system, the lower energy state ($\vect{y}_{e1}$) corresponds to the in-phase symmetric mode while the upper energy state ($\vect{y}_{e2}$) corresponds to the out-of-phase anti-symmetric mode \cite{F03,R03}. The energy levels of the eigenmodes are independent of the history function $\boldsymbol{\phi}(t)$ (or simply the initial conditions if $\tau=0$); however, their populations (as quantified by the total power density) depend on the history. 

This is illustrated clearly in Fig. \ref{TLSFig}, which shows the normalized total power density for $m_1=m_2=1$, $\gamma=0.001$, $\kappa=0.002$, $k_1=1-\delta$, and $k_2=1+\delta$. \LG{The  detuning} $\delta$ is varied from \RV{-0.01 to 0.01} with various constant history functions as listed in Table \ref{T1}. Two other constant history functions ($\vect{y}_{e1} \, \& \, \vect{y}_{e2}$) corresponding to the lower and upper eigenstates, respectively, are considered as well. Also, the states are shown for the cases with no delay ($\tau=0$) in the upper panels and for $\tau=0.1$ in the lower panels. Further details of the effects of the delay are shown in the results section.

\begin{table} [ht]
\caption{\label{T1} The considered constant history functions in Fig. \ref{TLSFig}.}
\begin{ruledtabular}
\begin{tabular}{l|cccc}
 & $y_1$ & $y_2$ & $y_3$ & $y_4$ \\
\hline
$\vect{y}_1$ & 1 & 0 & 0 & 0 \\
$\vect{y}_2$ & 0 & 0 & 1 & 0 \\
$\vect{y}_+$ & $1/\sqrt{2}$ & 0 & $1/\sqrt{2}$ & 0 \\
$\vect{y}_-$ & $1/\sqrt{2}$ & 0 & $-1/\sqrt{2}$ & 0
\end{tabular}
\end{ruledtabular}
\end{table}

\begin{figure}[t]
\centering
\includegraphics[width=3.4in]{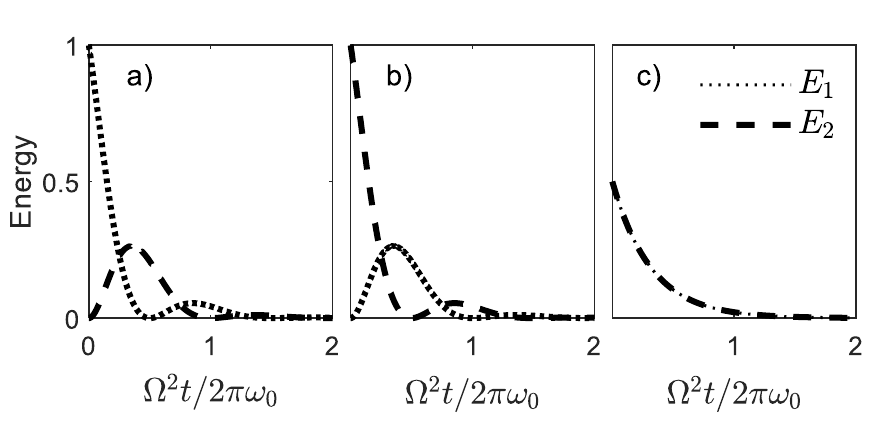}
\caption{The dynamics of oscillators total energies for three different constant history functions; namely a) $\vect{y}_1$, b) $\vect{y}_2$, and c) $\vect{y}_{e1}$, where $m_1=m_2=m=1$, $\gamma=0.001$, $\kappa=0.002$, $k_1=k_2=k=1$, $\tau=0$, $\omega_0^2=k/m$, and $\Omega^2=\kappa/m$. a) $\vect{y}_1$, b) $\vect{y}_2$, and c) $\vect{y}_{e1}$.}
\label{TLSDyn}
\end{figure}
    
\LG{The total energy of each oscillator \cite{F03,R03} is used to monitor the dynamics of individual energy states in the population}  i.e. $E_j(t)=T_j(t)+V_j(t)=\frac{m_j}{2} \dot x^2(t) + \frac{k_j+\kappa}{2} x^2(t)$, where $T_j$ and $V_j$ are the time-averaged kinetic and potential energies of the $j^\text{th}$ oscillator. Fig. \ref{TLSDyn} shows $E_1(t)$ and $E_2(t)$ for three different constant history functions; namely $\vect{y}_1$, $\vect{y}_2$, and $\vect{y}_{e1}$, where $m_1=m_2=m=1$, $\gamma=0.001$, $\kappa=0.002$, $k_1=k_2=k=1$, and $\tau=0$. Obviously, there is an energy transfer for mixed states while for the pure eigenstate, the population decays exponentially with no energy exchange. In this section, our objective is to illustrate how the dynamics is calculated for instantaneous interaction. Cases with delays are presented in the following results section.

\begin{figure}[t]
\centering
\includegraphics[width=3.4in]{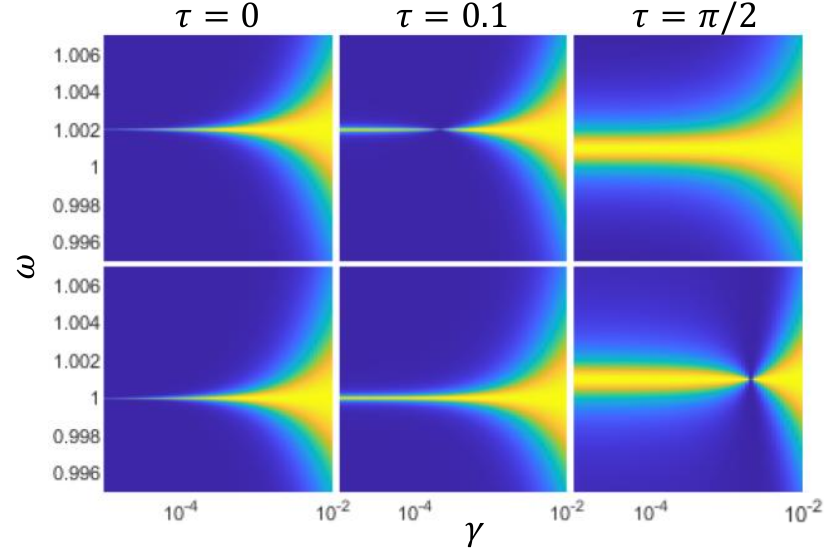}
\caption{The line wdiths of $\vect{y}_{e1}$ (lower panels) \& $\vect{y}_{e2}$ (upper panels) vs. the damping rate $\gamma$ when $\tau=0, 0.1, \& \, \pi/2$.}
\label{LWidth1}
\end{figure}

In the cases of $\vect{y}_{1},\, \vect{y}_{2},\, \vect{y}_{+},\, \& \, \vect{y}_{-}$, the states are mixed between the two eigenstates, \LG{, resulting in a distributed total power spectral density}; thus, \LG{as illustrated in Fig. \ref{TLSDyn}(a,b), energy exchange occurs.} In the cases of $\vect{y}_{e1}\, \& \, \vect{y}_{e2}$, the constant history functions \LG{are} pure eigenstates \LG{which precludes} energy exchange. \LG{Thus, as shown in Fig. \ref{TLSDyn}c,} the dynamics is simply a direct exponential decay of the initially populated eigenstate.

\LG{Additionally, the eigenstates clearly have a linewidth.} This is, of course, proportional to the damping rates $\gamma_j$, as illustrated in Fig \ref{LWidth1}. However, \LG{as shown in the figure}, the delay has a nonlinear effect on the linewidth, i.e., \LG{it affects dephasing. This is explored in more detail} in the results section.

\subsection{Energy Transfer and Coherence Quantification}

Energy exchange between different states can be characterized by two main ``measures'', energy exchange rate ($W_{i\to j}$) and coherence time  ($T_\text{Coh}$). $W_{i\to j}$ is the net instantaneous power transferred from the $i^\text{th}$ oscillator to the $j^\text{th}$ oscillator \cite{Z01} due to their delayed coupling by $\kappa$, i.e. the difference between the work per time unit due to coupling. \LG{Due to the fact that} the power density is concentrated around the oscillators' normal frequencies, the coupling results in a splitting that is \LG{proportional to $\kappa$ and a function of $\tau$}.  $W_{i\to j}$ is found to be oscillatory with a frequency \LG{equal to}  the difference between the altered normal frequencies of the oscillators due to the coupling and the delay. Thus, energy exchange dynamics can be quantified by this frequency difference,
\begin{equation}
    \label{FreqDiff}
    \omega_\text{Exc} = | \Im \left[{\lambda_i-\lambda_j} \right] | \, .
\end{equation}

In this work, \LG{we quantify coherence using} the cross-correlation $G_{12}(t)$ between the \LG{dynamics of two oscillators} and their full width at half maximum (FWHM) \cite{L04,C02,B01}, where
\begin{equation}
    \label{CrsCrl}
    G_{12}(t) = \int_{-\infty}^{\infty} E_1(s) E_2(s+t) ds \,.
\end{equation}
$G_{12}(t)$ is a measure of the similarity of temporal behaviors of the two oscillators' energies. For highly correlated quantities, $G_{12}(t)$ lives longer when compared to uncorrelated dynamics. The extension of $G_{12}(t)$ in term of FWHM ($T_\text{Coh}$) is hence a good measure to assess the coherence.

\section{\label{SecResults}Results \& Discussion}

In this work, we will consider an identical coupled system with normalized parameters for simpler presentation and discussion. \LG{This is sufficient for the purpose of this paper's analysis. However, more realistic values and alternative systems can be used while the majority of qualitative approaches remain unchanged.} Here, $m_1=m_2=m=1$ \& $k_1=k_2=k=1$. $\kappa$ is presented as a function of $\gamma=\gamma_1=\gamma_2$, which is set to 0.01. Thus, the normalized frequency $\omega_0$ is 1. The \LG{results and the subsequent analyses} show \LG{clearly} that the effects depend on the induced phase ($\omega_0 \tau$).

\begin{figure}[ht!]
\centering
\includegraphics[width=3.5in]{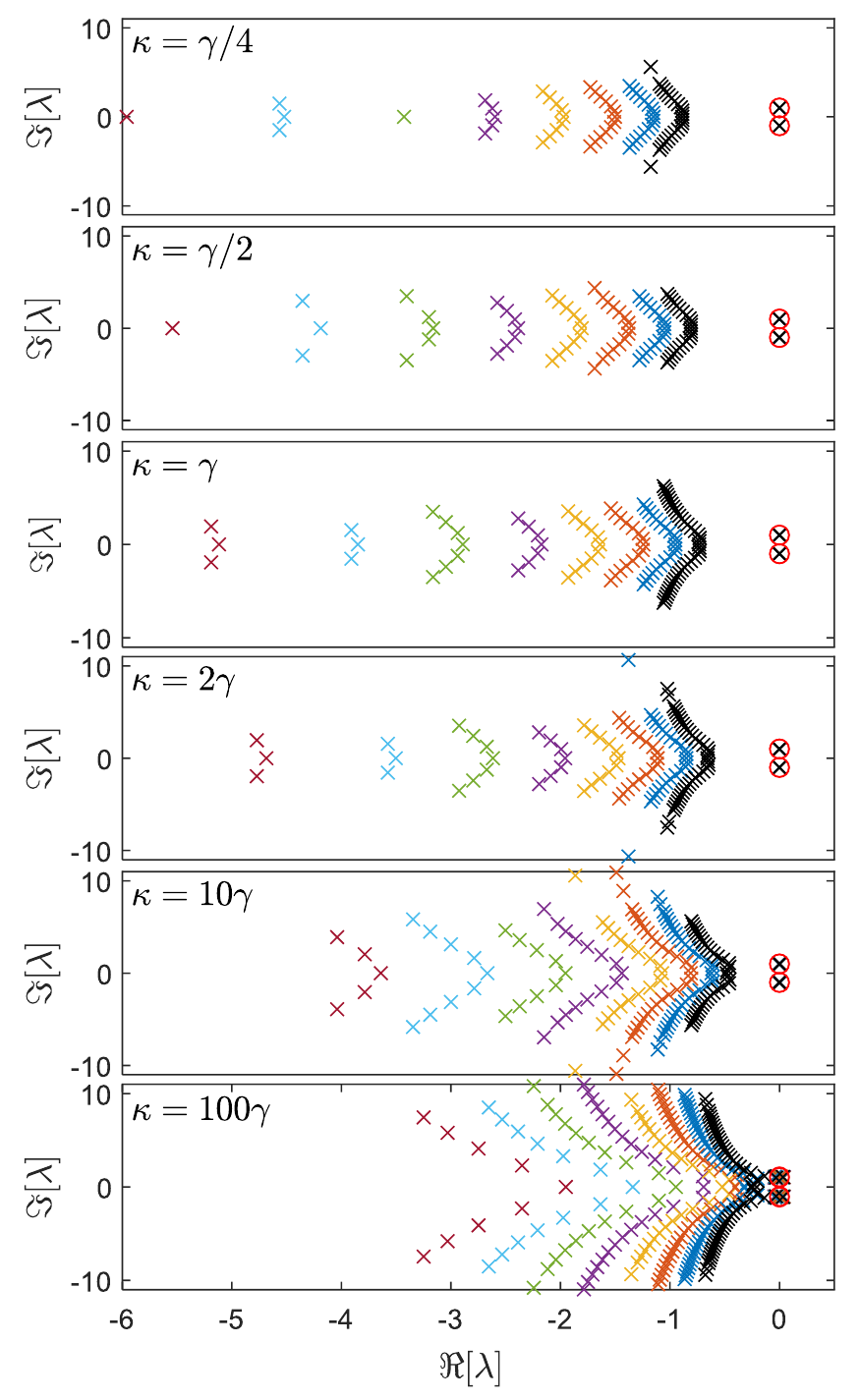}
\caption{The eigenvalue spectra of various $\kappa$'s for $\tau$ ranged between 0.3 and 1.0 at 0.1 steps where $m=1$, $\gamma=0.01$, and $k=1$. Colors codes: $\tau=1.0$ Black, $\tau=0.9$ blue, $\tau=0.8$ red, $\tau=0.7$ orange, $\tau=0.6$ purple, $\tau=0.5$ green, $\tau=0.4$ cyan, and $\tau=0.3$ maron. The red circles are for the main dynamical eigenstates.}
\label{Fig5}
\end{figure}

\begin{figure}[ht!]
\centering
\includegraphics[width=3.5in]{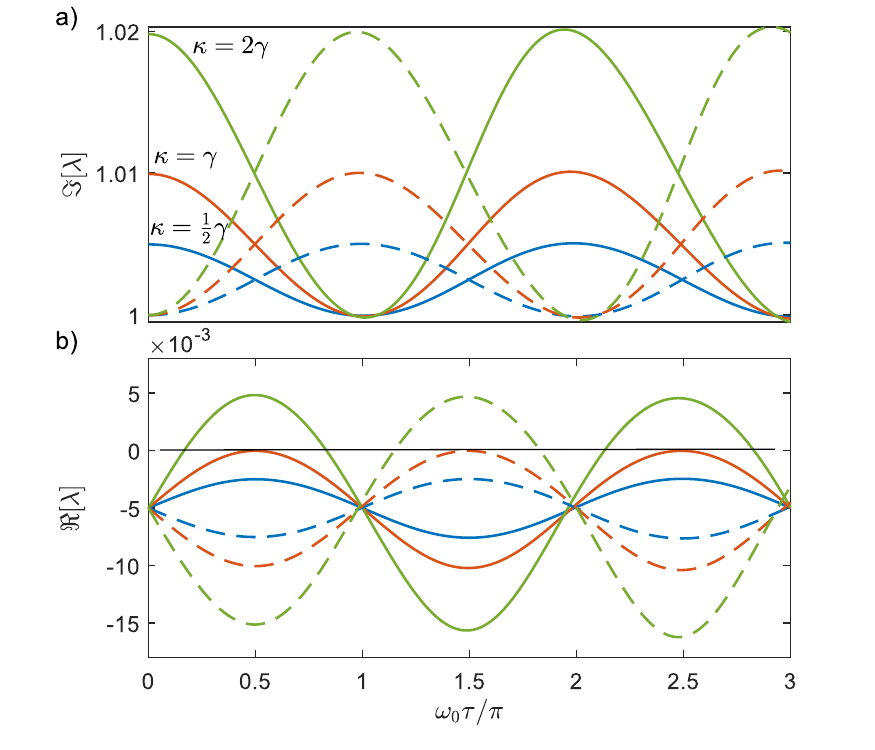}
\caption{The real and imaginary part of the main dynamical eigenvalues (around the instantaneous interaction eigenvalues) vs. $\tau$ for $\kappa=2\gamma, \, \gamma, \, \& \, \gamma/2$. The solid and dashed lines of the same color correspond to the two main dynamical eigenvalues for a particular $\kappa$. The black solid line is the line of dynamical stability when the real part becomes 0.}
\label{Fig6}
\end{figure}

\subsection{Spectral Analysis: The Effect of the Delay on Dynamical Eigenspace}

As discussed in the previous section, \LG{the} sufficient condition for stability is $\kappa/m < \gamma/2$. To test this, in the first analysis, the eigenvalue spectra of various $\kappa$'s for $\tau$ ranged between 0.3 and 1.0 at 0.1 steps are calculated and plotted in the complex plane as shown in Fig. \ref{Fig5}. First, it is clear that the spacing between the eigenvalues decreases with $\tau$. \LG{Only} four finite eigenvalues (in the red circles) exist for $\tau=0$ while the remaining ones lie at $-\infty$. However, more are brought to the right with increasing $\tau$. \LG{It} is also clear that the eigenvalues are shifted more to the right with increasing $\kappa$. It is actually this second shift \LG{is the one} that destabilize the system. By itself, $\tau$ can't destabilize the system for any $\kappa/m < \gamma/2$. In fact, we found that the limit is \LG{actually} $\kappa/m < \gamma$ as can be seen in Fig. \ref{Fig6}b. Evidently, the real part of the eigenvalues for $\kappa/m = \gamma$ is bounded from the top by 0 (the black horizontal line); i.e. it is globally negative and hence the dynamics \LG{remains} stable regardless of $\tau$.

\begin{figure*}[ht!]
\centering
\includegraphics[width=7.0in]{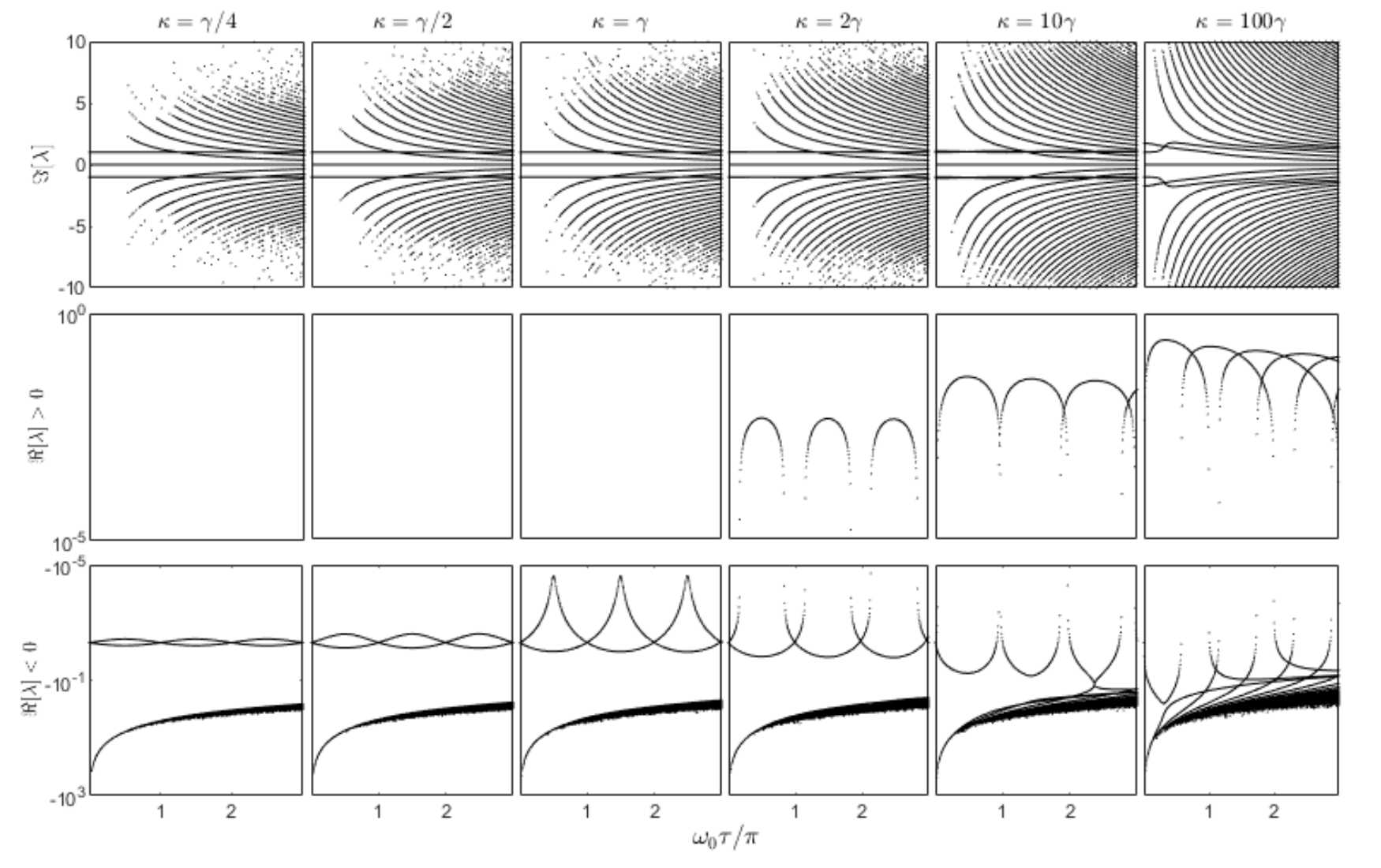}
\caption{The most right eigenvalues vs. $\omega_0\tau/\pi$ for various $\kappa$'s. The top row of panels are for the frequencies $\Im[\lambda]$, the middle row is for $\Re[\lambda]>0$ while the bottom row is for $\Re[\lambda]<0$.}
\label{Fig8}
\end{figure*}

Fig. \ref{Fig6} illustrates \LG{the oscillatory} effects of the delay on the main dynamical eigenvalues. This is expected in the considered harmonic system where the induced phase is $\omega_0 \tau$. The maximum deviation of the real parts of the main dynamical eigenvalues from the instantaneous interaction (i.e. $\gamma/2$) is $\kappa/2$ when $\omega_0 \tau=n \pi /2$ where $n$ is an odd integer \LG{and} the maximum splitting -- as expected -- is $\kappa$. In terms of energy, the maximum splitting between the two states also equals the coupling strength $\kappa$ and occurs for $\omega_0 \tau=n \pi$ where $n$ is an integer. Quantitatively, it is found (as shown in the following subsection) that the splittings in the real and imaginary parts are approximately $\kappa | \sin \left( \omega_0 \tau \right) |$ and $\kappa | \cos \left( \omega_0 \tau \right) |$. 

Since the splitting amplitude depends on $\kappa$ and oscillates \LG{for changing $\tau$}, $\kappa$ \LG{has an effect on the stability}. \LG{The} system remains stable for any $\kappa/m < \gamma$ regardless of $\tau$. Beyond this limit, the system is conditionally stable for certain \LG{intervals} of $\tau$.\LG{This  case is discussed in greater detail in the following subsection, which is devoted to stability.} For energy levels, the splitting becomes very small when $\omega_0 \tau=n \pi /2$ where $n$ is an odd integer and there seems to be a state crossing. However, we believe that there could be a manifestation of ``crossing avoidance''. Investigating this is beyond the scope of this paper and will be considered in a future work.

\subsection{Dynamics and stability}

In the second \LG{part} of the analysis, we investigate the effect of $\tau$ on the main dynamical modes and the stability. First, the eigenvalue spectra of various $\kappa$'s for $\omega_0\tau$ ranged between 0 and $3\pi$ \LG{are calculated and shown in Fig. \ref{Fig8} where $m=1$, $\gamma=0.001$, and $k=1$}. It can be seen that more eigenvalues \LG{shift to} the right with increasing $\tau$ and $\kappa$. However, none of them is asymptotically unstable (i.e. with $\Re[\lambda]>0$) for $\kappa\leq \gamma$ regardless of $\tau$. \LG{This shouldn't come as a surprise, since the delay only slows down communication rather than strengthen it.} Yet, it plays some stabilizing role for stronger coupling when $\kappa > \gamma$ as shown shortly.

\begin{figure}[ht!]
\centering
\includegraphics[width=3.3in]{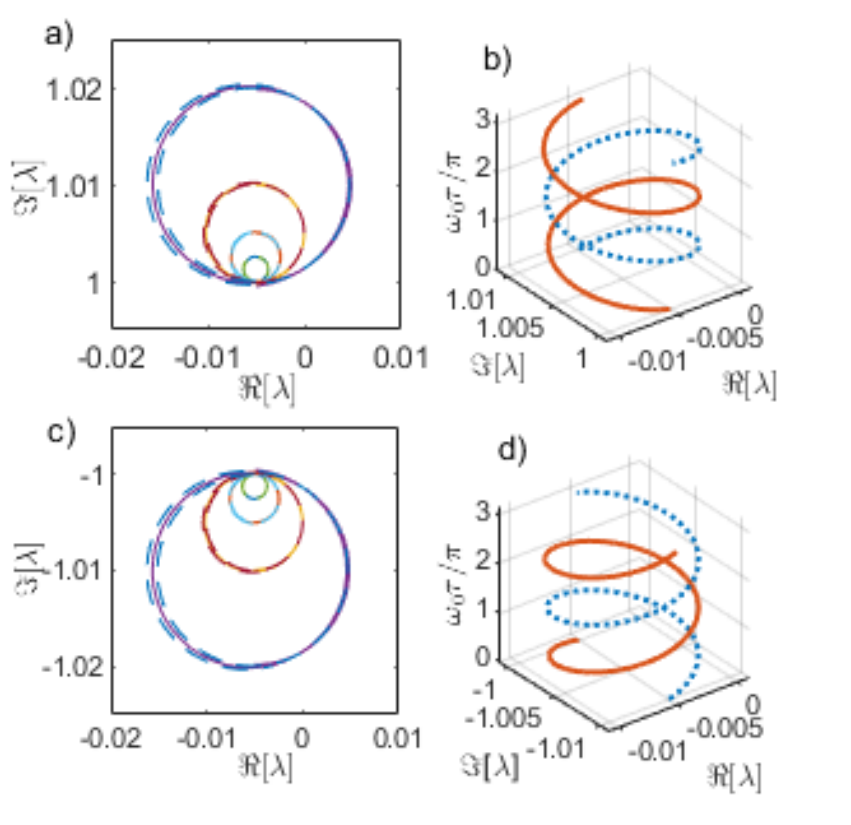}
\caption{The evolution of the main dynamical eigenvalues in the complex plane. a) and b) are for the upper coupled modes while c) and d) are for the lower conjugate modes.}
\label{Fig9}
\end{figure}

To understand this effect, we \LG{closely} analyze the changes of the main oscillators' eigenmodes (around the uncoupled ones) vs. $\tau$. The resulted evolutions of the two main modes are shown in Fig. \ref{Fig9}. It is found that the eigenvalues of the main modes reside on circles centered at $\left( -\frac{1}{2}\gamma \pm i \frac{k+\kappa}{m} \right)$ with a radius of $\kappa/2$. They are separated such that $|\lambda_1-\lambda_2|=\kappa$ as shown in Fig. \ref{Fig9}a and \ref{Fig9}c. When $\tau=0$, the two coupled modes and their conjugates reside \LG{on} a vertical line with $\Re[\lambda]=-\frac{1}{2}\gamma$. The upper two modes evolve and interwind clockwise (Fig. \ref{Fig9}b) around the upper circle with $\tau$ and they return back approximately to their original \LG{values}  when $\omega_0\tau=2n\pi$, where $n$ is an integer. The lower two modes evolve anticlockwise (Fig. \ref{Fig9}d) with $\tau$. It is found that the circles get more distorted with increasing $\kappa$.

\begin{figure}[ht!]
\centering
\includegraphics[width=2.0in]{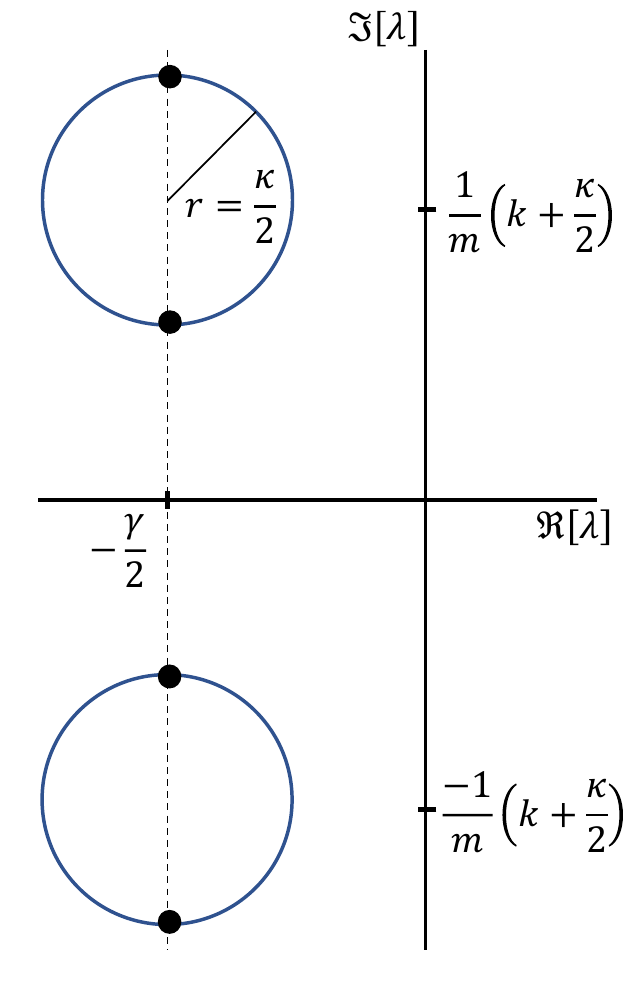}
\caption{A schematic of the evolution circles and interwinding of the main dynamical eigenmodes.}
\label{Fig7}
\end{figure}

\begin{figure}[ht!]
\centering
\includegraphics[width=3.2in]{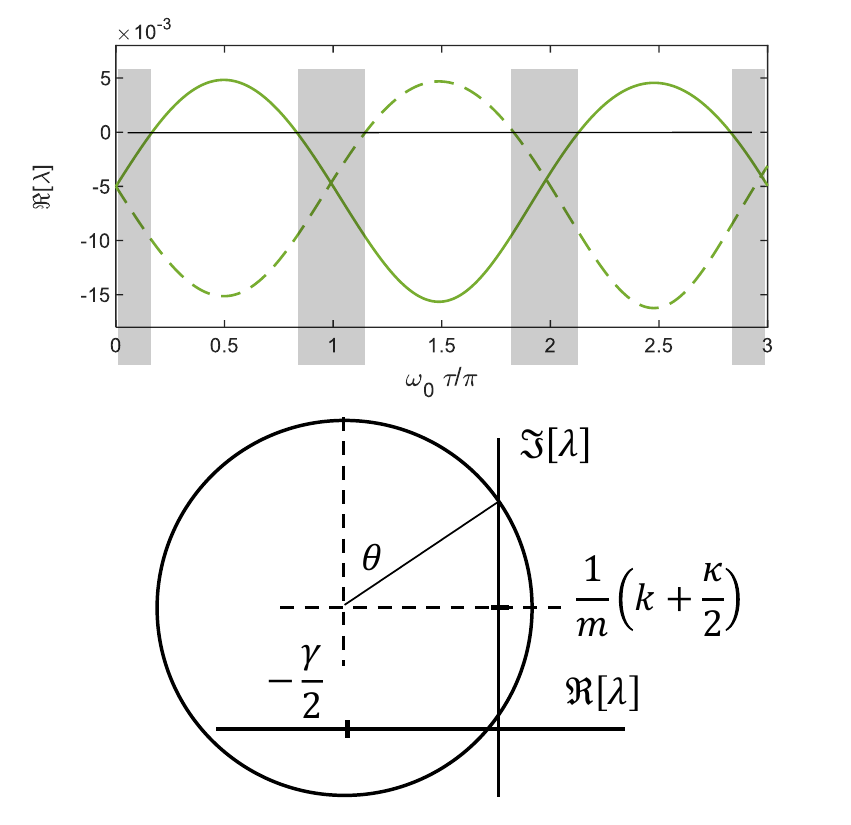}
\caption{Intervals of instability with increasing $\kappa$. The gray intervals in the upper panel correspond to stable dynamics.}
\label{Fig10}
\end{figure}

\begin{figure}[ht!]
\centering
\includegraphics[width=3.2in]{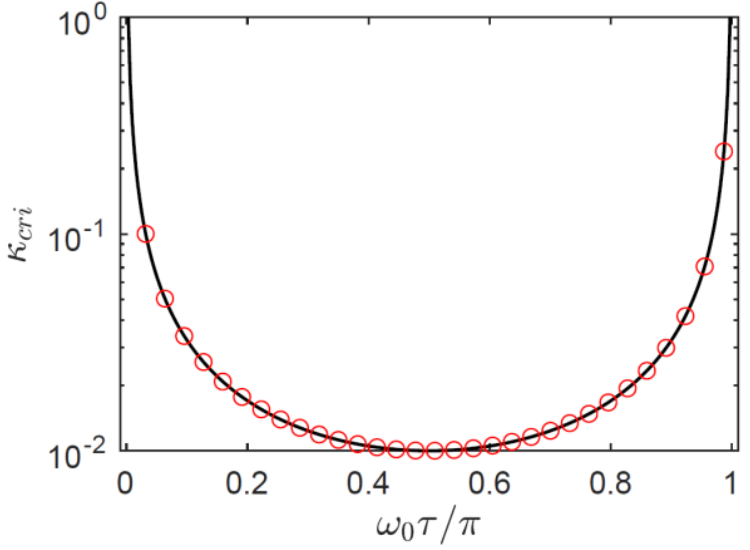}
\caption{\RV{The curve in black is for the bifurcation diagram of the studied system, where, the critical $\kappa$ is plotted vs. $\omega_0 \tau / \pi$. The red circles are according to the obtained stability condition in Eq. \ref{kappacri}.}}
\label{kvstau}
\end{figure}

\LG{This behavior of} evolution can be represented schematically by two evolution circles as shown in Fig. \ref{Fig7}. \LG{The} main two modes can \LG{then} be approximated \LG{as}:
\begin{equation}
\label{MEVal}
    \lambda_{1,2} \approx \left( -\frac{\gamma}{2} + i \frac{k+\kappa}{m} \right) \pm \frac{\kappa}{2m} \left( \sin \left( \omega_0 \tau \right) + i \cos \left( \omega_0 \tau \right) \right).
\end{equation}
\LG{Other conjugate modes can be handled in a similar fashion.} Eq. (\ref{MEVal}) implies that all the modes exist in the left complex plane if $\kappa<\gamma$. However, when $\kappa$ gets larger than $\gamma$, part of the evolution circle enters the right-side of the complex plane as shown in Fig. \ref{Fig10}. Yet, the main dynamical modes remain stable for some vertical \LG{bandwidth} (gray intervals in the upper panel of Fig. \ref{Fig10}). These strips are corresponding to:
\begin{equation}
\label{kappacri}
 n\pi-\sin^{-1}\left( \frac{\gamma}{\kappa} \right) < \omega_0\tau < n\pi+\sin^{-1}\left( \frac{\gamma}{\kappa} \right)
\end{equation}
where $n$ is an integer. \LG{These intervals} get thinner \LG{with} increasing $\kappa$. \RV{This is further verified by a numerical bifurcation analysis \cite{E01} using DDE-BIFTOOL \cite{E02} as shown in Fig. \ref{kvstau}. The calculated critical coupling $\kappa_\text{cri}$ by bifurcation matches those of Eq. \ref{kappacri}.}

\begin{figure}[ht!]
\centering
\includegraphics[width=3.4in]{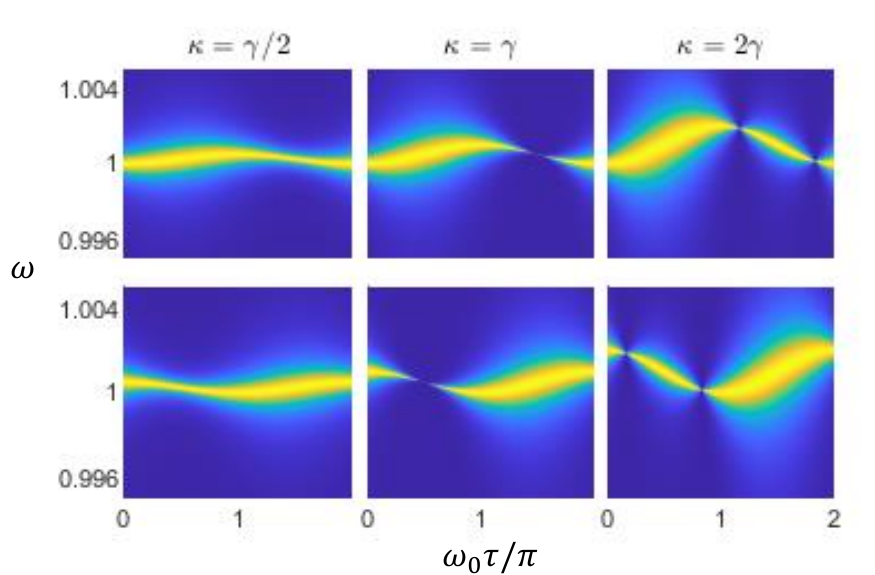}
\caption{The total power density $P_t(\omega)$ vs. $\tau$ for three different coupling case; $\kappa=\gamma/2$, $\kappa=\gamma$, and $\kappa=2\gamma$. The upper panels correspond to $\vect{y}_{e1}$ while the lower panels are for $\vect{y}_{e2}$.}
\label{Fig11}
\end{figure}

\subsection{Two-level system with delay: States, Energy Transfer, and Coherence Quantification}

\LG{Here we will investigate the impact of delay on main eigenmode power density.} Fig. \ref{Fig11} presents the power densities of the two eigenmodes $\vect{y}_{e1}$ (upper panels) and $\vect{y}_{e2}$ (lower panels) corresponding to in-phase and out-of-phase respectively for three different coupling cases; $\kappa=\gamma/2$, $\kappa=\gamma$, and $\kappa=2\gamma$. In this part, the delay's effect is \LG{evident}. \LG{The first primary effect of } $\tau$ \LG{is that it} tunes the splitting, although the maximum splitting is still $\kappa$ \LG{dependent}. This is \LG{manifested in} Eq. (\ref{MEVal}) where the frequencies are 
\begin{equation*}
    \Im \left[ \lambda_{1,2} \right] = \frac{k+\kappa}{m} \pm \frac{k}{2m} \cos \left( \omega_0 \tau \right) \, .
\end{equation*}
So obviously, this effect is periodic and gets repeated every $\omega_0\tau=2\pi$. 

\LG{$\tau$ also has a strong effect on the linewidth of the two level states.} As observed, at some spots, the linewidth can get infinitesimally small. This \LG{occurs} when one of the main modes crosses the imaginary \LG{axis. The other state's linewidth becomes large at this point.} \LG{Damping cancellation is the cause of this effect, which should have a significant impact on the likelihood of transitioning from one state to the other.}

\begin{figure}[ht!]
\centering
\includegraphics[width=3.5in]{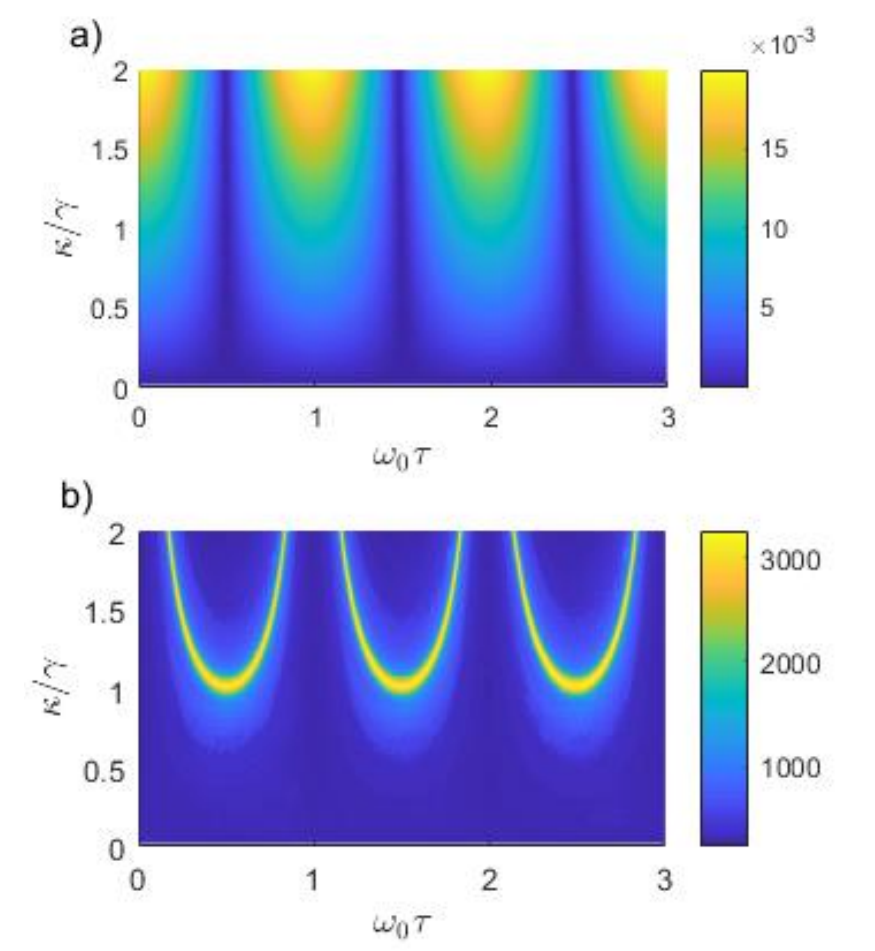}
\caption{a) The energy exchange frequency $w_\text{Exc}$ and b) coherence time $T_\text{Coh}$ vs. $\kappa$ and $\tau$. }
\label{Fig12}
\end{figure}

\begin{figure}[ht!]
\centering
\includegraphics[width=3.2in]{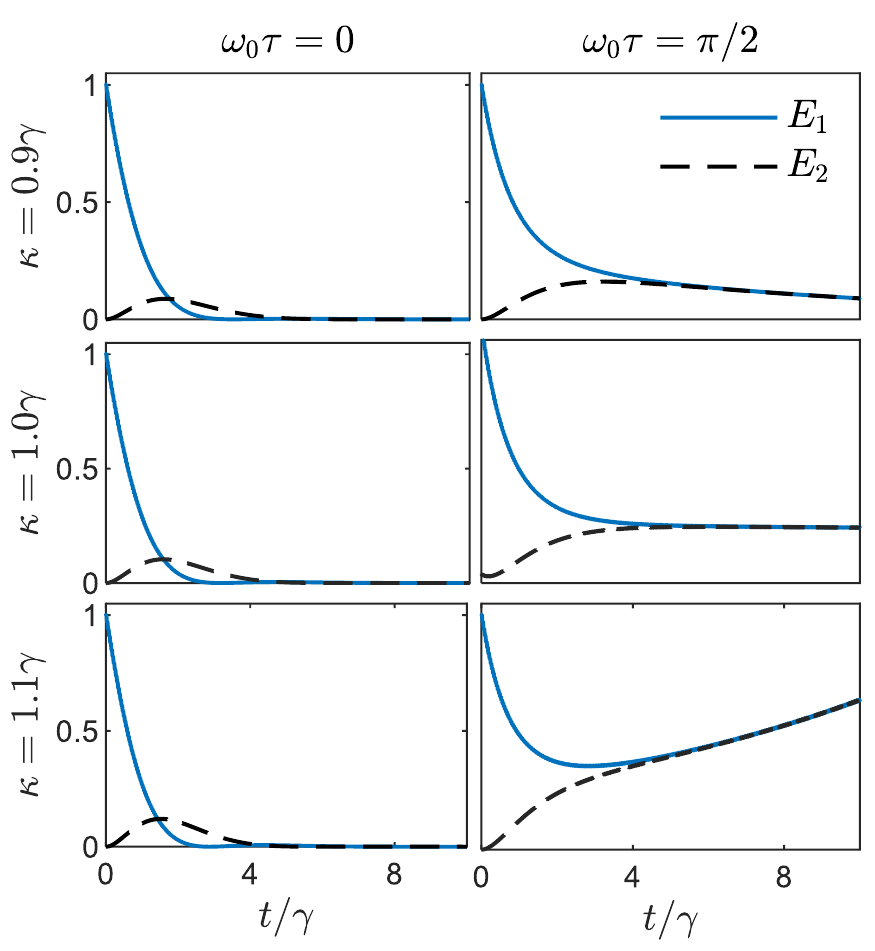}
\caption{The dynamics of the two oscillators' energies ($E_1$ and $E_2$) are calculated for $\kappa=0.9\gamma, \, 1.0\gamma, \, \& \, 1.1\gamma$ (in the three rows from the top) and for $\omega_0\tau=0 \, \& \, \pi/2$ (left and right panels respectively).}
\label{Fig13}
\end{figure}

\LG{In the following, we'll look into} energy transfer and their coherence between the two level system. In this analysis, $\kappa$ is varied between $0.02\gamma$ and $2\gamma$, and $\omega_0\tau$ is varied between 0 and $3\pi$. \LG{The} energy exchange frequency $w_\text{Exc}$ and coherence time $T_\text{Coh}$ are calculated and shown in Fig. \ref{Fig12}\LG{, where one can see that } $w_\text{Exc}$ (Fig. \ref{Fig12}a) increases with $\kappa$; but, it is considerably influenced by $\tau$ as well. \LG{The value of $w_\text{Exc}$} is maximum when the main dynamical eigenvalues \LG{lie} on a vertical line and \LG{is} minimum when they \LG{lie} on a horizontal one. In the \LG{latter} case and when $\kappa$ approaches $\gamma$, the real parts of the eigenmodes become $-\gamma$ and 0; i.e. one mode is rapidly damped while the second becomes undamped due to damping cancellation; \LG{a typical behavior of nonlinear systems} \cite{P02,K05}. It is this undamped mode that allows a good part of the energy to live longer as shown in Fig. \ref{Fig13}.

In Fig. \ref{Fig13}, the dynamics of the two oscillators energies ($E_1$ and $E_2$) are calculated for $\kappa=0.9\gamma, \, 1.0\gamma, \, \& \, 1.1\gamma$ (\LG{the upper} three rows) and for $\omega_0\tau=0 \, \& \, \pi/2$ (left and right panels respectively). For \LG{cases} with $\omega_0\tau=0$ (left panels), both $E_1$ and $E_2$ decay rapidly and almost vanish for $t=4/\gamma$. \LG{The case of $\omega_0\tau=\pi/2$, on the other hand, falls into one of three categories.}  $\omega_0\tau=\pi/2$:
\begin{itemize}
    \item When $\kappa=0.9\gamma$: the system is still stable with the real parts for the two modes equal to $-0.95\gamma$ and $-0.05\gamma$. Both are decaying; but the second one is dying \LG{slower} and hence $E_1$ and $E_2$ live longer.
    \item When $\kappa=1.0\gamma$: the system is still marginally stable with the real parts for the two modes equal to $-\gamma$ and $0$. The second mode becomes undamped and $E_1$ and $E_2$ should reach a steady state and remain constant after some time (no energy exchange afterward).
    \item When $\kappa=1.1\gamma$: The real parts for the two modes are $-1.05\gamma$ and $0.05\gamma$, respectively. Thus, the second mode becomes unstable and $E_1$ and $E_2$ will unphysically grow.
\end{itemize}
\LG{This behavior is evident in the obtained $T_\text{Coh}$, as illustrated in Fig. \ref{Fig12}b.} $T_\text{Coh}$ is enhanced significantly when one of the main dynamical modes lies on or close to the imaginary axis of the complex plane.

\begin{figure}[ht!]
\centering
\includegraphics[width=3.4in]{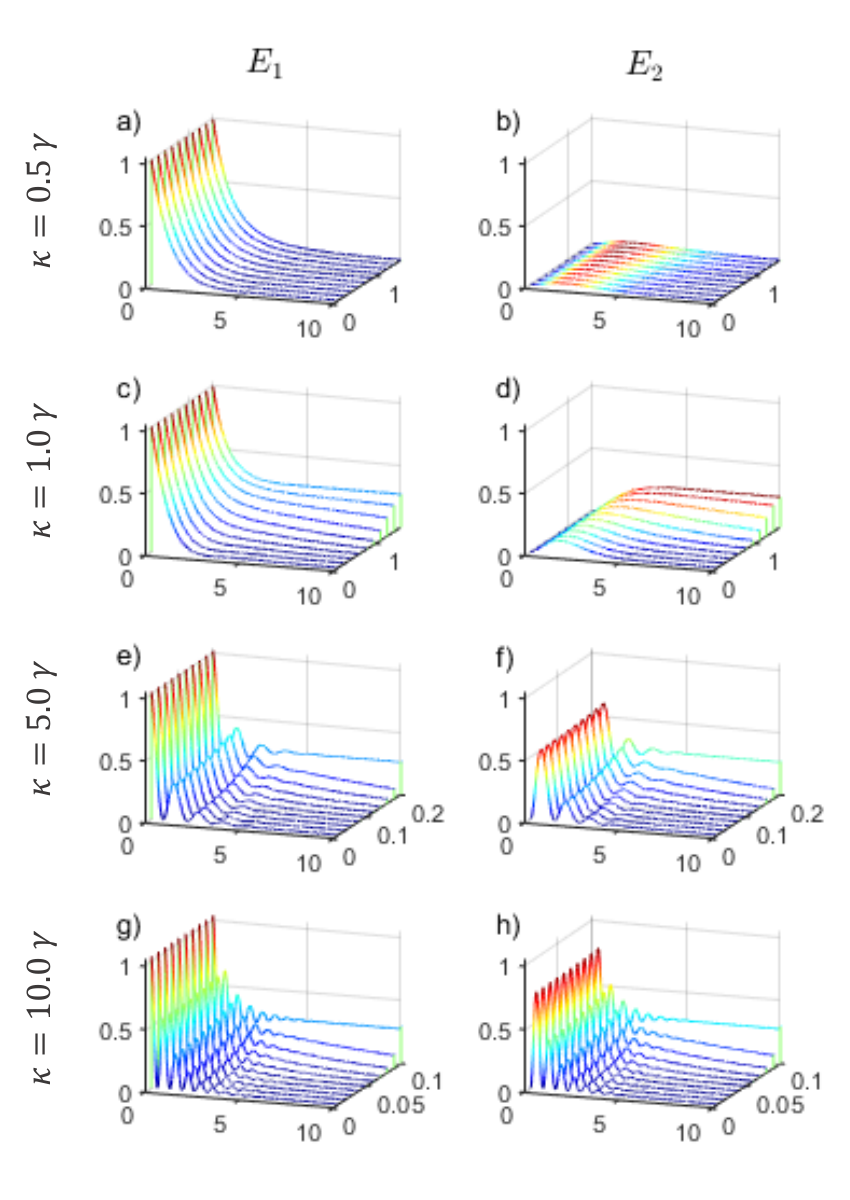}
\caption{The dynamics of the two oscillators' energies ($E_1$ and $E_2$) are calculated for $\kappa=0.5\gamma, \, 1\gamma,  \, 5\gamma, \, \& \, 10\gamma$ (in the four rows from the top) and for $\omega\tau=0 \, \& \, \pi/2$ (left and right panels respectively). In each panel, the $x$-axis represents the time from 0 till $\gamma t=10$, the $y$-axis is the delay ranges between 0 and $\omega_0 \tau_{\text{Cri}}$, and the $z$-axis is the energy.}
\label{Fig14}
\end{figure}

\LG{Next, we'll look at the case where the $\kappa$ and $\gamma$ values are almost equal.} In this range \LG{(See Fig. 14)}, when one of the modes crosses (or gets close to) the imaginary axis of the complex plane, $w_\text{Exc}$ becomes very small as shown previously in Fig. \ref{Fig10}, \LG{which implies} that energy exchange \LG{also becomes}  very small. \LG{In  practical cases, where $\kappa \gg \gamma$, the system should be unstable at this limit; however, it becomes stable for certain ranges of $\omega_0\tau$.} The first of these ranges starts from the case \LG{of} no delay ($\tau=0$) till $\tau_{\text{Cri}}=\frac{1}{\omega_0} \sin^{-1}(\gamma/\kappa)$. When a mode crosses the imaginary axis, $w_\text{Exc}=\kappa \sqrt{1-\frac{\gamma^2}{\kappa^2}}$, i.e. increases with $\kappa$ and equals to 0 when $\kappa=\gamma$. In Fig. \ref{Fig14}, $E_1(t)$ and $E_2(t)$ are shown for $\kappa/\gamma=0.5, \, 1, \, 5, \& \, 10$. The left panels correspond to $E_1(t)$ while the right panels correspond to $E_2(t)$. For the first case, $\kappa/\gamma=0.5$ and the energies decay very rapidly. In the second case, $\kappa/\gamma=1$. \LG{As the delay approaches $\tau_{\text{Cri}}$, one can see that the damping of one of the modes is canceled out when $w_\text{Exc}=0$.}  This means that no energy exchange occurs after some time and both $E_1(t)$ and $E_2(t)$ becomes constant. In the last two cases when $\kappa/\gamma=5$ and 10, at $\tau_{\text{Cri}}$, $w_\text{Exc}$ is still finite and hence an energy exchange exists and a coherence is maintained further for a \LG{longer} time.

\section{\label{SecCon}Conclusions}

In this work, the Frimmer-Novotny model \cite{F03,F05,R03} to \LG{emulate} two-level systems is extended by incorporating a constant time delay in the coupling and studying its effects on system dynamics and the two-level modeling. Mathematically, the \LG{inclusion of the time} delay converts the dynamical system from a finite  one \LG{to} an infinite-dimensional system. \LG{Due to the Hookean nature of the coupling, the problem can be solved analytically as a linear autonomous system of delay differential equations.} \LG{The Krylov method, with Chebyshev interpolation and post-processing refinement, is used to solve the mathematical model.}

The model is used to study the dynamics of the two-level system with delayed interaction. \LG{A delay turns out to be a critical factor, as revealed by the calculations and analyses.} Its effects are oscillatory as the main dynamical eigenmodes \LG{evolve} around a circle with a radius proportional to the coupling strength and an angle proportional to $\tau$. This oscillation governs the energy transfer dynamics and coherence. The effects are diverse and substantial; but, one of the most important effects happens when one of the main modes crosses the imaginary axis of the complex plane, where it become purely imaginary and dampingless. Thus, the two states energies can live and be exchanged for an extremely long time. \LG{The delay is also found to have an impact on} both splitting and the linewidth (a manifestation of damping cancellation). This influences the energy transfer and the coherence further.

The studied model can \LG{become} unstable when \LG{the coupling is stromg, i.e.} $\kappa>\gamma$. However, for certain intervals of the delay $\tau$, the system \LG{remains} stable even for $\kappa \gg \gamma$. \LG{For vanishing $\tau$}, the system is globally stable and remains stable until \LG{$\tau_\text{Cri}=\frac{1}{\omega_0} \sin^{-1}(\gamma/\kappa)$ is reached, at which time} the system becomes dampingless and the energies can live and be exchanged for an extremely long \LG{period of} time. \LG{The realistic delay should not be excessively large; it should also be observable in real-world systems. For example, critical tau can be measured in tens of attoseconds for an optical system with a 500 nm wavelength.}

\section{Aknowledgement}

This is the Accepted Manuscript version of an article accepted for publication in Physica Scripta. IOP Publishing Ltd is not responsible for any errors or omissions in this version of the manuscript or any version derived from it. This Accepted Manuscript is published under a CC BY licence. The Version of Record is available online at DOI: 10.1088/1402-4896/ac7f62.

\nocite{*}

\bibliography{CODelay}

\end{document}